%%      JNL.TEX                                 Doug Eardley
 %%
 \message
 {JNL.TEX version 0.92 as of 6/9/87.  Report bugs and problems to Doug Eardley.}
 %%
 %%      This is a set of TeX 82 macros designed to produce scientific
 %%      papers with a minimum of fuss and using as much of plain.tex as
 %%      possible.  The user need only know what is in the TeXbook, and
 %%      the macros under ``user definitions'' below.  Also, the user
 %%      definitions are intended to be as simple as possible, so that
 %%      the user may change them as desired.  I have tried to avoid all
 %%      cleverness, although it may have snuck in here and there.
 %%
 %%      A considerable degree of compatibility with AmSTeX is maintained,
 %%      although not guaranteed.  The intention is that AmSTeX input file
 %%      should run with only a few changes near the beginning;  see
 %%      discussion below under "AmSTeX compatability".
 %%
 %%      For documentation, see the file JNLHLP.TEX.  Optional features are
 %%      contained in the files PPT.TEX (for two-up preprints), REFORDER.TEX
 %%      (automatic numbering of references), EQNORDER.TEX (automatic numbering
 %%      of equations), and TABLEOFC.TEC (automatic generation of table of
 %%      contents).
 
 %%
 %%      Redefine \input to prevent files being loaded more than once
 %%
 \catcode`@=11
 \expandafter\ifx\csname inp@t\endcsname\relax\let\inp@t=\input
 \def\input#1 {\expandafter\ifx\csname #1IsLoaded\endcsname\relax
 \inp@t#1%
 \expandafter\def\csname #1IsLoaded\endcsname{(#1 was previously loaded)}
 \else\message{\csname #1IsLoaded\endcsname}\fi}\fi
 \catcode`@=12

 %%
 %%  Font definitions suitable for the IMAGEN-480 (Written by Tony Kennedy)
 %%
 
 %  Define a whole menagerie of pseudo-12pt fonts
 
 \font\twelverm=cmr10 scaled 1200    \font\twelvei=cmmi10 scaled 1200
 \font\twelvesy=cmsy10 scaled 1200   \font\twelveex=cmex10 scaled 1200
 \font\twelvebf=cmbx10 scaled 1200   \font\twelvesl=cmsl10 scaled 1200
 \font\twelvett=cmtt10 scaled 1200   \font\twelveit=cmti10 scaled 1200
 \font\twelvesc=cmcsc10 scaled 1200  \font\twelvesf=cmssdc10 scaled 1200
 %\skewchar\twelvei='177   \skewchar\twelvesy='60
 
 %  Define \...point macros to change fonts and spacings consistently
 
 \def\twelvepoint{\normalbaselineskip=12.4pt plus 0.1pt minus 0.1pt
   \abovedisplayskip 12.4pt plus 3pt minus 9pt
   \abovedisplayshortskip 0pt plus 3pt
   \belowdisplayshortskip 7.2pt plus 3pt minus 4pt
   \smallskipamount=3.6pt plus1.2pt minus1.2pt
   \medskipamount=7.2pt plus2.4pt minus2.4pt
   \bigskipamount=14.4pt plus4.8pt minus4.8pt
   \def\rm{\fam0\twelverm}          \def\it{\fam\itfam\twelveit}%
   \def\sl{\fam\slfam\twelvesl}     \def\bf{\fam\bffam\twelvebf}%
   \def\mit{\fam 1}                 \def\cal{\fam 2}%
   \def\sc{\twelvesc}               \def\tt{\twelvett}
   \def\sf{\twelvesf}
   \textfont0=\twelverm   \scriptfont0=\tenrm   \scriptscriptfont0=\sevenrm
   \textfont1=\twelvei    \scriptfont1=\teni    \scriptscriptfont1=\seveni
   \textfont2=\twelvesy   \scriptfont2=\tensy   \scriptscriptfont2=\sevensy
   \textfont3=\twelveex   \scriptfont3=\twelveex  \scriptscriptfont3=\twelveex
   \textfont\itfam=\twelveit
   \textfont\slfam=\twelvesl
   \textfont\bffam=\twelvebf \scriptfont\bffam=\tenbf
   \scriptscriptfont\bffam=\sevenbf
   \normalbaselines\rm}
 
 %       tenpoint

 %%
 %%      Various internal macros
 %%
 
 \def\beginlinemode{\endmode
   \begingroup\parskip=0pt \obeylines\def\\{\par}\def\endmode{\par\endgroup}}
 \def\beginparmode{\endmode
   \begingroup \def\endmode{\par\endgroup}}
 \let\endmode=\par
 {\obeylines\gdef\
 {}}
 \def\singlespace{\baselineskip=\normalbaselineskip}
 
 \def\oneandahalfspace{\baselineskip=\normalbaselineskip
   \multiply\baselineskip by 3 \divide\baselineskip by 2}
 \def\doublespace{\baselineskip=\normalbaselineskip \multiply\baselineskip by 2}

 \newcount\firstpageno
 \firstpageno=2
 \footline={\ifnum\pageno<\firstpageno{\hfil}\else{\hfil\twelverm\folio
  \hfil}\fi}
 \def\toppageno{\global\footline={\hfil}\global\headline
   ={\ifnum\pageno<\firstpageno{\hfil}\else{\hfil\twelverm\folio\hfil}
  \fi}}
 \let\rawfootnote=\footnote              % We must set the footnote style
 \def\footnote#1#2{{\rm\singlespace\parindent=0pt\parskip=0pt
   \rawfootnote{#1}{#2\hfill\vrule height 0pt depth 6pt width 0pt}}}
 \def\raggedcenter{\leftskip=4em plus 12em \rightskip=\leftskip
   \parindent=0pt \parfillskip=0pt \spaceskip=.3333em \xspaceskip=.5em
   \pretolerance=9999 \tolerance=9999
   \hyphenpenalty=9999 \exhyphenpenalty=9999 }
 \def\dateline{\rightline{\ifcase\month\or
   January\or February\or March\or April\or May\or June\or
   July\or August\or September\or October\or November\or December\fi
   \space\number\year}}
 \def\received{\vskip 3pt plus 0.2fill
  \centerline{\sl (Received\space\ifcase\month\or
   January\or February\or March\or April\or May\or June\or
   July\or August\or September\or October\or November\or December\fi
   \qquad, \number\year)}}
 
 %%
 %%      Page layout, margins, font and spacing (feel free to change)
 %%
 
 \hsize=6.5truein
 %\hoffset=1truein
 \hoffset=0pt
 \vsize=8.9truein
 %\voffset=1truein
 \voffset=0pt
 \parskip=\medskipamount
 \def\\{\cr}
 \twelvepoint            % selects twelvepoint fonts (cf. \tenpoint)
 \doublespace            % selects double spacing for main part of paper (cf.
                         %       \singlespace, \oneandahalfspace)
 \overfullrule=0pt       % delete the nasty little black boxes for overfull box
 
 %%
 %%      The user definitions for major parts of a paper (feel free to change)
 %%
 
   %  "Draft", Timestamp
 
   % Preprint number at upper right of title page
 
 \def\title                      %  Title on title page
   {\null\vskip 3pt plus 0.2fill
    \beginlinemode \doublespace \raggedcenter \bf}
 
 \def\author                     %  Author(s) name(s)  on title page
   {\vskip 3pt plus 0.2fill \beginlinemode
    \singlespace \raggedcenter\sl}    %%% atention j'ai chang~N sc -> sl
 
 \def\affil                      % Affiliations (can intermix with \author)
   {\vskip 3pt plus 0.1fill \beginlinemode
    \oneandahalfspace \raggedcenter \sl}
 
 \def\abstract                   % Begin abstract
   {\vskip 3pt plus 0.3fill \beginparmode
    \oneandahalfspace ABSTRACT: }
 
 \def\endtitlepage               % End title page, begin body of paper
   {\endpage                     %       This subsumes \body
    \body}
 \let\endtopmatter=\endtitlepage
 
 \def\body                       % Begin text body;  can be used to end
   {\beginparmode}               % \title, \author, \affil, \abstract,
                                 % \reference, or \figurecaption modes
 
 \def\head#1{                    % Head;  NOTE enclose the text in {}
   \goodbreak\vskip 0.5truein    %  e.g., \head{I. Introduction}
   {\immediate\write16{#1}
    \raggedcenter \uppercase{#1}\par}
    \nobreak\vskip 0.25truein\nobreak}

 \def\beginitems{
 \par\medskip\bgroup\def\i##1 {\item{##1}}\def\ii##1 {\itemitem{##1}}
 \leftskip=36pt\parskip=0pt}
 \def\enditems{\par\egroup}
 
 \def\beneathrel#1\under#2{\mathrel{\mathop{#2}\limits_{#1}}}

 \def\refto#1{[#1]}
 
 \def\references                 % Begin references -- basic format is Phys Rev
   {\head{References}            % I.e., volume, page, year (space after commas).
    \beginparmode
    \frenchspacing \parindent=0pt \leftskip=1truecm
    \parskip=8pt plus 3pt \everypar{\hangindent=\parindent}}

 \gdef\refis#1{\item{#1.\ }}                     % Ref list numbers.

 \gdef\journal#1, #2, #3, 1#4#5#6{           % Journal reference. Comma sets
 {\sl #1~}{\bf #2}, #3 (1#4#5#6)}            % off: name, vol, page, year

                            % off: name, year, vol, page.

 \def\endreferences{\body}
 
 \def\figurecaptions             % Begin figure captions
   {\endpage
    \beginparmode
    \head{Figure Captions}
 }
 
 \def\endfigurecaptions{\body}
 
 \def\endpage                    %  Eject a page
   {\vfill\eject}
 
 \def\endpaper                   %  Ways to say goodbye
   {\endmode\vfill\supereject}

 %%
 %%      AmSTeX compatability definitions
 %%
 %%      To run a TeX file originally intended for AmSTeX, only small changes
 %%      should be necessary (I hope).  Use the line \input jnl at the start.
 %%      Remove the lines \input amstex, \documentstyle{itpjnl} at the
 %%      beginning;  also remove all the page layout stuff (\parindent=1cm,
 %%      \hsize=5.28125in etc.)  The page layout is now done automatically.
 %%      Also OMIT the qualifier \magnification=1200 when you IMPRINT the
 %%      .dvi file.  (\TagsOnRight is harmless, you can take it out or leave
 %%      it in.)  I believe most AmSTeX will work with no change.  One problem
 %%      is \footnote, which is a little different in that it now needs to
 %%      have an explicit asterisk *  (or whatever) included, like this:
 %%              \footnote*{Text winds up at bottom of page.}
 %%      This is discussed on p. 116 of the TeXbook.  IGNORE the AmSTeX
 %%      documentation (if you can call it that);  refer to the TeXbook.
 %%
 %%      Note that many commands in AmSTeX have their equivalents in the
 %%      TeXbook, perhaps with different names and slightly differing
 %%      usage. E.g., the old \align in AmSTeX is replaced by \eqalign
 %%      (p. 190) and \aligntag is replaced by \eqalignno (p. 192).
 %%      \align and \aligntag still work, but I recommend that you use
 %%      \eqalign and \eqalignno in documents run under jnl.
 %%
 %%      See me if you have any problems  -- Doug.
 %%

 \def\heading                            % Heading
   {\vskip 0.5truein plus 0.1truein      % e.g., \heading I. NOTES \endheading
    \beginparmode \def\\{\par} \parskip=0pt \singlespace \raggedcenter}

 \def\subheading                         % Subheading
   {\vskip 0.25truein plus 0.1truein     % e.g., \subheading{A. The Problem}
    \beginlinemode \singlespace \parskip=0pt \def\\{\par}\raggedcenter}

 \def\tag#1$${\eqno(#1)$$}
 
 \def\align#1$${\eqalign{#1}$$}

 \def\aligntag#1$${\gdef\tag##1\\{&(##1)\cr}\eqalignno{#1\\}$$
   \gdef\tag##1$${\eqno(##1)$$}}
 
 \def\endaligntag{}

 \def\overset #1\to#2{{\mathop{#2}\limits^{#1}}}
 \def\underset#1\to#2{{\let\next=#1\mathpalette\undersetpalette#2}}
 \def\undersetpalette#1#2{\vtop{\baselineskip0pt
 \ialign{$\mathsurround=0pt #1\hfil##\hfil$\crcr#2\crcr\next\crcr}}}

 %%
 %%      Various little user definitions
 %%
 
 \def\ref#1{Ref.~#1}                     %       for inline references
 \def\Ref#1{Ref.~#1}                     %       ditto
 \def\[#1]{[\cite{#1}]}
 \def\cite#1{{#1}}
                    % For figure numbers
          % For citation of equation numbers
        %       ditto
                     %       ditto
                   %       ditto
                 %       ditto
 \def\(#1){(\call{#1})}
 \def\call#1{{#1}}
 \def\taghead#1{}
 \def\frac#1#2{{#1 \over #2}}

 \def\12{{1\over2}}

 \def\sla{\raise.15ex\hbox{$/$}\kern-.57em}
 \def\leaderfill{\leaders\hbox to 1em{\hss.\hss}\hfill}
 \def\twiddle{\lower.9ex\rlap{$\kern-.1em\scriptstyle\sim$}}
 \def\bigtwiddle{\lower1.ex\rlap{$\sim$}}
 \def\gtwid{\mathrel{\raise.3ex\hbox{$>$\kern-.75em\lower1ex\hbox{$\sim$}}}}
 \def\ltwid{\mathrel{\raise.3ex\hbox{$<$\kern-.75em\lower1ex\hbox{$\sim$}}}}
 \def\square{\kern1pt\vbox{\hrule height 1.2pt\hbox{\vrule width 1.2pt\hskip 3pt
    \vbox{\vskip 6pt}\hskip 3pt\vrule width 0.6pt}\hrule height 0.6pt}\kern1pt}
 \def\tdot#1{\mathord{\mathop{#1}\limits^{\kern2pt\ldots}}}
 
 \def\pmb#1{\setbox0=\hbox{#1}%
   \kern-.025em\copy0\kern-\wd0
   \kern  .05em\copy0\kern-\wd0
   \kern-.025em\raise.0433em\box0 }

 \catcode`@=11
 \newcount\r@fcount \r@fcount=0
 \newcount\r@fcurr
 \immediate\newwrite\reffile
 \newif\ifr@ffile\r@ffilefalse
 \def\w@rnwrite#1{\ifr@ffile\immediate\write\reffile{#1}\fi\message{#1}}
 
 \def\writer@f#1>>{}
 \def\referencefile{%                      Stuff to write .REF file
   \r@ffiletrue\immediate\openout\reffile=\jobname.ref%
   \def\writer@f##1>>{\ifr@ffile\immediate\write\reffile%
     {\noexpand\refis{##1} = \csname r@fnum##1\endcsname = %
      \expandafter\expandafter\expandafter\strip@t\expandafter%
      \meaning\csname r@ftext\csname r@fnum##1\endcsname\endcsname}\fi}%
   \def\strip@t##1>>{}}

 \def\citeall#1{\xdef#1##1{#1{\noexpand\cite{##1}}}}
 \def\cite#1{\each@rg\citer@nge{#1}}     % Variable No. of args, separated by ","
 
 \def\each@rg#1#2{{\let\thecsname=#1\expandafter\first@rg#2,\end,}}
 \def\first@rg#1,{\thecsname{#1}\apply@rg}       % each@ag is a general purpose
 \def\apply@rg#1,{\ifx\end#1\let\next=\relax%      variable no. of arg. macro.
 \else,\thecsname{#1}\let\next=\apply@rg\fi\next}% args separated by commas
 
 \def\citer@nge#1{\citedor@nge#1-\end-}  % Check for M-N range (M and N numbers)
 \def\citer@ngeat#1\end-{#1}
 \def\citedor@nge#1-#2-{\ifx\end#2\r@featspace#1 % Single argument
   \else\citel@@p{#1}{#2}\citer@ngeat\fi}        % M-N range of arguments
 \def\citel@@p#1#2{\ifnum#1>#2{\errmessage{Reference range #1-#2\space is bad}.%
     \errhelp{If you cite a series of references by the notation M-N, then M and
     N must be integers, and N must be greater than or equal to M.}}\else
  {\count0=#1\count1=#2\advance\count1 by1\relax\expandafter\r@fcite\the\count0,%
   \loop\advance\count0 by1\relax%         Loop from M to N
     \ifnum\count0<\count1,\expandafter\r@fcite\the\count0,%
   \repeat}\fi}
 
 \def\r@featspace#1#2 {\r@fcite#1#2,}    % Eat spaces at beginning or end of arg
 \def\r@fcite#1,{\ifuncit@d{#1}%           Cite individual reference
     \newr@f{#1}%
     \expandafter\gdef\csname r@ftext\number\r@fcount\endcsname%
                      {\message{Reference #1 to be supplied.}%
                       \writer@f#1>>#1 to be supplied.\par}%
  \fi%
  \csname r@fnum#1\endcsname}
 \def\ifuncit@d#1{\expandafter\ifx\csname r@fnum#1\endcsname\relax}%
 \def\newr@f#1{\global\advance\r@fcount by1%
     \expandafter\xdef\csname r@fnum#1\endcsname{\number\r@fcount}}
 
 \let\r@fis=\refis                       % Save old \refis, redefine
 \def\refis#1#2#3\par{\ifuncit@d{#1}%      Use two params #2 #3 to strip
 blank
    \newr@f{#1}%
    \w@rnwrite{Reference #1=\number\r@fcount\space is not cited up to now.}\fi%
   \expandafter\gdef\csname r@ftext\csname r@fnum#1\endcsname\endcsname%
   {\writer@f#1>>#2#3\par}}
 
 \def\ignoreuncited{%   redefine \refis if ignoring uncited references
    \def\refis##1##2##3\par{\ifuncit@d{##1}%
      \else\expandafter\gdef\csname r@ftext\csname r@fnum##1\endcsname\endcsname%
      {\writer@f##1>>##2##3\par}\fi}}
 
 \def\r@ferr{\endreferences\errmessage{I was expecting to see
 \noexpand\endreferences before now;  I have inserted it here.}}
 \let\r@ferences=\references
 \def\references{\r@ferences\def\endmode{\r@ferr\par\endgroup}}
 
 \let\endr@ferences=\endreferences
 \def\endreferences{\r@fcurr=0%            Save old \endreferences, redefine
   {\loop\ifnum\r@fcurr<\r@fcount%         Loop over refnum and produce text
     \advance\r@fcurr by 1\relax\expandafter\r@fis\expandafter{\number\r@fcurr}%
     \csname r@ftext\number\r@fcurr\endcsname%
   \repeat}\gdef\r@ferr{}\endr@ferences}
 
 % Save old \endpaper, redefine it to write parting message.
 
 \let\r@fend=\endpaper\gdef\endpaper{\ifr@ffile
 \immediate\write16{Cross References written on []\jobname.REF.}\fi\r@fend}
 
 \catcode`@=12

 \citeall\refto          % These macros will generate citations
 \citeall\ref            %
 \citeall\Ref            %

\catcode`@=11
\newcount\tagnumber\tagnumber=0

\immediate\newwrite\eqnfile
\newif\if@qnfile\@qnfilefalse
\def\write@qn#1{}
\def\writenew@qn#1{}
\def\w@rnwrite#1{\write@qn{#1}\message{#1}}
\def\@rrwrite#1{\write@qn{#1}\errmessage{#1}}

\def\taghead#1{\gdef\t@ghead{#1}\global\tagnumber=0}
\def\t@ghead{}

\expandafter\def\csname @qnnum-3\endcsname
  {{\t@ghead\advance\tagnumber by -3\relax\number\tagnumber}}
\expandafter\def\csname @qnnum-2\endcsname
  {{\t@ghead\advance\tagnumber by -2\relax\number\tagnumber}}
\expandafter\def\csname @qnnum-1\endcsname
  {{\t@ghead\advance\tagnumber by -1\relax\number\tagnumber}}
\expandafter\def\csname @qnnum0\endcsname
  {\t@ghead\number\tagnumber}
\expandafter\def\csname @qnnum+1\endcsname
  {{\t@ghead\advance\tagnumber by 1\relax\number\tagnumber}}
\expandafter\def\csname @qnnum+2\endcsname
  {{\t@ghead\advance\tagnumber by 2\relax\number\tagnumber}}
\expandafter\def\csname @qnnum+3\endcsname
  {{\t@ghead\advance\tagnumber by 3\relax\number\tagnumber}}

\def\equationfile{%
  \@qnfiletrue\immediate\openout\eqnfile=\jobname.eqn%
  \def\write@qn##1{\if@qnfile\immediate\write\eqnfile{##1}\fi}
  \def\writenew@qn##1{\if@qnfile\immediate\write\eqnfile
    {\noexpand\tag{##1} = (\t@ghead\number\tagnumber)}\fi}
}

\def\callall#1{\xdef#1##1{#1{\noexpand\call{##1}}}}
\def\call#1{\each@rg\callr@nge{#1}}

\def\each@rg#1#2{{\let\thecsname=#1\expandafter\first@rg#2,\end,}}
\def\first@rg#1,{\thecsname{#1}\apply@rg}
\def\apply@rg#1,{\ifx\end#1\let\next=\relax%
\else,\thecsname{#1}\let\next=\apply@rg\fi\next}

\def\callr@nge#1{\calldor@nge#1-\end-}
\def\callr@ngeat#1\end-{#1}
\def\calldor@nge#1-#2-{\ifx\end#2\@qneatspace#1 %
  \else\calll@@p{#1}{#2}\callr@ngeat\fi}
\def\calll@@p#1#2{\ifnum#1>#2{\@rrwrite{Equation range #1-#2\space is bad.}
\errhelp{If you call a series of equations by the notation M-N, then M and
N must be integers, and N must be greater than or equal to M.}}\else%
 {\count0=#1\count1=#2\advance\count1 by1\relax\expandafter\@qncall\the\count0,%
  \loop\advance\count0 by1\relax%
    \ifnum\count0<\count1,\expandafter\@qncall\the\count0,%
  \repeat}\fi}

\def\@qneatspace#1#2 {\@qncall#1#2,}
\def\@qncall#1,{\ifunc@lled{#1}{\def\next{#1}\ifx\next\empty\else
  \w@rnwrite{Equation number \noexpand\(>>#1<<) has not been defined yet.}
  >>#1<<\fi}\else\csname @qnnum#1\endcsname\fi}

\let\eqnono=\eqno
\def\eqno(#1){\tag#1}
\def\tag#1$${\eqnono(\displayt@g#1 )$$}

\def\aligntag#1\endaligntag
  $${\gdef\tag##1\\{&(##1 )\cr}\eqalignno{#1\\}$$
  \gdef\tag##1$${\eqnono(\displayt@g##1 )$$}}

\def\eqalignno#1{\displ@y \tabskip\centering
  \halign to\displaywidth{\hfil$\displaystyle{##}$\tabskip\z@skip
    &$\displaystyle{{}##}$\hfil\tabskip\centering
    &\llap{$\displayt@gpar##$}\tabskip\z@skip\crcr
    #1\crcr}}

\def\displayt@gpar(#1){(\displayt@g#1 )}

\def\displayt@g#1 {\rm\ifunc@lled{#1}\global\advance\tagnumber by1
        {\def\next{#1}\ifx\next\empty\else\expandafter
        \xdef\csname @qnnum#1\endcsname{\t@ghead\number\tagnumber}\fi}%
  \writenew@qn{#1}\t@ghead\number\tagnumber\else
        {\edef\next{\t@ghead\number\tagnumber}%
        \expandafter\ifx\csname @qnnum#1\endcsname\next\else
        \w@rnwrite{Equation \noexpand\tag{#1} is a duplicate number.}\fi}%
  \csname @qnnum#1\endcsname\fi}

\def\ifunc@lled#1{\expandafter\ifx\csname @qnnum#1\endcsname\relax}

\let\@qnend=\end\gdef\end{\if@qnfile
\immediate\write16{Equation numbers written on []\jobname.EQN.}\fi\@qnend}

\catcode`@=12

%% DEBUG
%%\def\see#1 {\expandafter\show\csname#1\endcsname}

%%
%%
%%
%%
%%    MAIN TEXT OF THE PAPER 
%%
%%
%%
%%
%%

\def\lsim{\lower.8ex\hbox{$\buildrel<\over\sim$}}
\def\gsim{\lower.8ex\hbox{$\buildrel>\over\sim$}}
\def\chap{\mathaccent 94}

\def\dbar#1{\underline{\underline{#1}}}
\def\til{\mathaccent "7E }

\def\lim#1 {\displaystyle{\mathop{\ell im}_{#1}}}

\def\somd#1 {\displaystyle{\mathop{\sum \sum}_{#1} }}
\def\bra{\langle}
\def\ket{\rangle}
\def\gcall#1{(\call{#1})}

\def\bR{ {\bf R} }

\def\bV#1{ {\bf V}_{#1} }
\def\bv#1{ {\bf v_{#1}}}
\def\bP{{\bf P}}

\def\bx{ \chap{\bf x} }
\def\by{ \chap{\bf y} }
\def\bz{ \chap{\bf z} }

\def\sig1{ {\chap{\bf \sigma} } }

\def\Ff {{\bf\delta {\cal F}}}

\def\Fbar {{\bf \bar{\cal F}}}
\def\F0 {{{\bf\cal F}_0}}
\def\Fd0 {{{\bf\cal F}_0^\dagger}}
\def\r1{ {\chap{\bf r} } }
\def\ZTA#1 {\dbar{\zeta}_{#1}}
\def\ZTAN#1 {\dbar{\zeta}^N_{#1}}
\def\ZTAinf#1 {\dbar{\zeta}^{N=\infty}_{#1} }
\def\Unity {{\dbar {\bf 1}}}
\def\lsim{\lower.8ex\hbox{$\buildrel<\over\sim$}}
\def\gsim{\lower.8ex\hbox{$\buildrel>\over\sim$}}

\def\real {\hbox{I}\kern-0.1667em \hbox{R}}
\def\PP{ {\cal P} }
\def\FF{ {\Ff}}
\def\MM{ {\dbar{\cal M} } }
\def\UU{ {\dbar{\cal U} } }

\def\Gam{ {\dbar{\Gamma} } }

\head{\bf  BINARY FRICTION TENSOR FOR BROWNIAN PARTICLES: OVERCOMING SPURIOUS 
 FINITE-SIZE EFFECTS.}

\vskip 15pt
\author Lyd\'eric Bocquet, Jean-Pierre Hansen and Jaroslaw Piasecki (*)

\vskip 20pt

\affil{ Laboratoire de Physique, Ecole Normale Sup\'erieure
de Lyon (URA CNRS 1325), 
46 All\'ee d'Italie, 69007  Lyon (France)}

\affil{(*) Permanent address : Institute of Theoretical Physics, Warsaw University,
Hoza 69, 00-681 Warsaw (Poland)} 

\smallskip
\vskip 1cm

PACS numbers: 05.20.Dd, 05.40.+j, 82.70.Dd

\abstract
Starting from  a careful analysis of the coupled Langevin equations for two 
interacting Brownian particles, we derive a method for extracting the binary 
friction tensor from the correlation function matrix of the instantaneous 
forces exerted by the bath particles on the fixed Brownian particles, and from 
the relaxation of the total momentum of the bath in a {\it finite} system.
The general methodology, which circumvents the pitfalls associated with the 
inversion of the thermodynamic and long time limits, is applied to the case of 
two Brownian hard spheres in a bath of light spheres.
\vskip 40pt   
submitted to Physical Review Letters (May 96)
%\dateline

\endtopmatter

Ever since Green-Kubo (GK) formulae have been derived, expressing linear transport 
coefficients as time integrals of correlation functions of thermally fluctuating 
dynamical variables, it has been known that particular care must be exercised 
in evaluating these integrals from correlation functions for {\it finite} systems 
\refto{Kirkwood}. Strictly speaking, GK formulae yield non-zero results 
only provided the thermodynamic limit is taken {\it before} the upper limit 
in the GK integral is taken to infinity. To obtain sensible, non-zero results 
from the integration of correlation functions of {\it finite} systems, like 
those provided by Molecular Dynamics (MD) simulations of samples of 
$N\simeq 10^2 - 10^4$ particles, a somewhat arbitrary upper cut-off $\tau_{N}$ 
must be applied. For most transport coefficients, involving systems of 
identical or similar particles, the resulting values are not very sensitive to 
the precise value of $\tau_{N}$, since it is found that after a time of the order 
of the initial, fast relaxation of the system under study (typically a 
picosecond for dense fluids), the integral reaches a "plateau" value, which roughly 
coincides with the time beyond which the MD-generated correlation function 
drops below the noise level.

However, the difficulty is less easily overcome when one considers the classic 
example of the friction coefficient $\zeta$ exerted on a heavy Brownian particle
by a bath of much lighter particles. $\zeta$ is related to the time integral of 
the autocorrelation function (ACF) of the instantaneous force exerted by the bath 
particles on the Brownian particle. Recent MD simulations clearly show
that no well-defined ``plateau'' value of the GK integrand is observed in 
systems invoving several hundred bath particles, so that the cut-off time becomes
totally arbitrary \refto{Espanol,Bocquet}. In practice $\zeta$ was determined
from the relaxation of the total momentum of the fluid, due to the collisions
with a fixed Brownian particle of infinite mass $M$. In this letter, we will
consider the case of {\it two} Brownian particles suspended in a bath of 
discrete light particles. We will show that finite size effects are even 
more subtle in this system and lead to spurious results for the computed friction 
tensor. We will then present a method to overcome these effects and obtain the
correct friction tensor. This is a first step in a statistical (first 
principles) approach to the hydrodynamic interactions in suspensions; such
interactions are traditionnally derived from macroscopic hydrodynamics
\refto{Beenaker} which ignore the discrete nature of the bath, and are hence
expected to fail at nanometric scales, like those explored by modern surface
force machines \refto{Israel}.

The friction tensor $\ZTA{} $ relates the fluctuations of the forces acting on 
each suspended Brownian particle to the velocities of these particles :

$$
\eqalign{
\Ff_a(t) &= -\ZTA{a1} \bV{1} (t) - \ZTA{a2} \bV{2} (t); \ \ a=1,2 \cr
}
\eqno(rel_FV)
$$
where the $2\times 2$ matrix $\ZTA{ab} $ of friction tensors is a function
of the relative position $\bR=\bR_1-\bR_2$ of the two Brownian particles.
The tensor occurs naturally in the two-particle Fokker-Planck equation,
describing the dynamical evolution of two massive particles in a bath of
much lighter particles \refto{Mazo,Murphy,Deutch}. The resulting
microscopic  expression for the friction tensor reads :
$$
\ZTA{ab} = {1 \over {k_B T}} \int_0^\infty d\tau\ \bra \Ff(\bR_{a};0) \cdot
\Ff (\bR_{b};\tau) \ket_{(eq \vert \bR_{1},\bR_{2})}\ \
\eqno(def_zeta)
$$
where the notation $(eq \vert \bR_{1},\bR_{2})$ refers to an equilibrium
average over the fluid variables in the presence of two \underbar{fixed}
Brownian particles located at $\bR_1$ and $\bR_2$;
$
\Ff(\bR_{a};t)={\bf {\cal F}}(\bR_{a};t)-\bra {\bf {\cal F}}(\bR_{a}) 
\ket_{(eq \vert \bR_{1},\bR_{2})}
%\eqno(dF)
$
is the fluctuation of the force experienced by the Brownian particle at 
$\bR_{a}$ due to collisions with fluid particles. Due to the presence of the
other Brownian particle, the average force experienced by particle $a$ does
not vanish, and is substracted from the instantaneous force,
to yield the fluctuating force for a given configuration of the Brownian
particles. Recently we have provided a rigorous derivation of the two-particle
Fokker-Planck equation for a system of two Brownian spheres of diameter
$\Sigma$ in a fluid of $N$ smaller spheres of diameter $\sigma$ \refto{Piasecki}.
Use of the multiple time-scale analysis, previously applied to derive the Fokker-Planck
equation for a single Brownian particle \refto{Cukier, Bocquet}, avoids any
``ad hoc'' assumptions concerning the separation of time scales. For hard spheres,
the fluctuating force reduces to the rate of transfer of momentum from the bath
to Brownian particles in the course of instantaneous elastic collisions, as one
might have intuitively expected \refto{Alley}. The average force, on the other hand,
which will henceforth be denoted by $\Fbar_a$ for brevity, may be identified with the
familiar entropic depletion force acting between sterically stabilized
colloidal particles \refto{Asakura}.

The difficulties encountered in attempting to evaluate the friction tensor
from eq. \gcall{def_zeta} for a finite system are immediately apparent
if one notes that for a system of $N$ fluid particles in the presence of
two Brownian particles, the time derivative of the total fluid momentum
may be identified with the sum of the forces acting on the two Brownian
particles. It follows from this observation that for a finite system :
$$
\eqalign{
\ZTA{1a} + \ZTA{2a} &= {1 \over {k_B T}} \int_0^\infty d\tau\ \bra 
\left[\Ff(\bR_{1};\tau)+\Ff(\bR_{2};\tau)\right] \cdot
\Ff (\bR_{a};0) \ket_{(eq \vert \bR_{1},\bR_{2})}\ \ \cr
&= {1 \over {k_B T}} \int_0^\infty d\tau\ \bra 
\dot\bP \cdot
\Ff (\bR_{a};0) \ket_{(eq \vert \bR_{1},\bR_{2})}\ = 0 \cr
}
\eqno(som=0)
$$
where $\bP$ is the total fluid momentum.
Since there is no reason why the diagonal and off-diagonal (or mutual) friction
tensors should be exactly opposite, the spurious result \gcall{som=0} must
be regarded as a consequence of the finite size of the system.

In order to see how sensible results may be extracted from dynamical trajectories
of finite systems, we consider the coupled evolution equations of the momenta
$\bP_a$ of the two Brownian particles, in the form of generalized Langevin
equations, namely :
$$
\eqalign{
\dot{\bP_a} &=  \Fbar_a(t) + \sum_{b=1,2} \int_0^t d\tau\ \dbar{M}_{ab} (t-\tau) \cdot
\bP_b(\tau) \  
+ \Ff_a^{+} (t) \cr
}
\eqno(langevin)
$$
where $\dbar{M}_{ab}$ is the matrix of memory functions, and $\Ff_a^{+}(t)$ is the
fluctuating ``random''force, associated with the ``fast''fluid variables; it
satisfies the following constraints :
$$
\bra \Ff_a^{+}(t)\ket = 0,\ 
\bra \Ff_a^{+}(t) \bP_b(t^\prime) \ket = 0, \ 
\bra \Ff_a^{+}(t) \Fbar_b(t^\prime) \ket = 0 
\eqno(proprietes)
$$
Note that the mean forces $\Fbar_b$ depend on time through the slowly
varying Brownian particle positions. The clear separation of time scales
associated  with the heavy Brownian particles (slow variables), and with 
the light fluid particles (fast variables), justifies the usual Markovian
assumption, namely
$$
\bra \Ff_a^{+} (t) \Ff_b^{+} (t^\prime) \ket = 2 \Gam_{a,b}\ \delta (t-t^\prime)
\equiv 2 \left\{ \int_0^\infty d\tau\ \bra \Ff_a^{+} (\tau) \Ff_b^{+} (0) \ket
\right\}\ \delta (t-t^\prime) 
\eqno(Gam)
$$
which amounts to neglecting memory effects, so that the coupled generalized
Langevin equations \gcall{langevin} reduce to the phenomenological local
Langevin equations. Using matrix notations (e.g. $\PP=\left({\bP_1, \bP_2}\right)$ etc.),
they can be cast in the form
$$
\dot\PP (t) = \Fbar(t) + \til\MM \cdot \PP (t) \ + \ \Ff^{+}(t)
\eqno(langevin_mat)
$$
where the t-independent matrix $\til\MM$ is related to the ACF matrix 
of the random force via the fluctuation-dissipation theorem :
$$
\eqalign{
\til\MM &= \Gam \cdot \bra \PP\PP\ket^{-1} \cr
}
\eqno(rel_MM)
$$
Taking the Laplace transform of the Langevin eq. \gcall{langevin_mat}, and 
projecting onto $\FF(t=0)$ leads to the following relation for the 
fluctuating force ACF :
$$
\bra \til\FF (s)  \FF\ket = \left(\Unity + {\til\MM \over {\ s}} \right)^{-1}
\cdot \bra \til\Ff^{+} (s=0)  \Ff^{+} (0) \ket\equiv \UU^{-1}(s)
\bra \til\Ff^{+} (s=0)  \Ff^{+} (0) \ket
\eqno(equality)
$$
where the tilda denotes a Laplace transform and $s$ is the variable conjugate
to time.

The inverse of the correlation matrix of the Brownian particle momenta may be
calculated in the microcanonical ensemble (appropriate for MD simulations)
for a system of $2$ Brownian particles of mass $M$ and $N$ bath particles
of mass $m$, using the results of ref. \refto{Espanol} :
$$
\bra \PP  \PP \ket^{-1} = {1 \over{Mk_BT}}
\left(\matrix{
 {1-\lambda \over {1-2\lambda}} \Unity & {\lambda \over {1-2\lambda}} \Unity\cr
{\lambda \over {1-2\lambda}} \Unity    & { 1-\lambda \over {1-2\lambda}} \Unity\cr
}\right)
\eqno(IPPPP)
$$
where $\lambda = {M \over {2M+Nm}}$. If eq. \gcall{IPPPP} is substituted into
eq. \gcall{equality} and the thermodynamic limit is taken \underbar{before}
the Brownian limit $M\rightarrow \infty$, the matrix $\MM$ is found to vanish,
and comparison of relation \gcall{equality} taken for $s=0$, with the
GK relation \gcall{def_zeta} yields the following expression for the friction tensor
matrix :
$$
\ZTA{}  = 
{1 \over {k_BT}} \int_0^\infty \bra \Ff^{+}(0) \Ff^{+} (\tau) \ket \ d\tau 
\eqno(ZTA+)
$$
In other words, provided the limits are taken in the order specified
above, the friction tensor is given equivalently in terms of the ``bare''
or ``random'' force ACF. For a finite system, the two estimates, say
$\ZTA{} ^{N}$ and $\ZTA{} ^{N,+}$ corresponding respectively to
eqs. \gcall{def_zeta} and \gcall{ZTA+} differ. However, the spurious behaviour of
the bare force ACF, embodied in the result \gcall{som=0} which is a direct 
consequence of the conservation of the total momentum of a finite system,
is not expected to carry  over to the random (or projected) force ACF,
which are associated with the {\it fast} fluid variables. Consequently no singularity
is expected  when the order of limits ($N\rightarrow \infty,M\rightarrow
\infty$) is inverted, and hence the finite and infinite system results for 
$\ZTA{} ^{N,+}$ should only differ by terms of order $1/N$, i.e. :
$$
\ZTAinf{} =\ZTA{} ^{N,+} + {\cal O} \left( {1\over N}\right)
\eqno(diff)
$$
Discarding henceforth the superscript $N$ for the tensor $\ZTA{} ^{+}$ to simplify
notations, we first note that symmetry considerations imply that the tensors 
$\ZTA{ab} ^{+}$ are diagonal and
$$
\ZTA{11} ^{+} =\ZTA{22} ^{+} =
\zeta_{s\perp}^{+} \left(\bx \bx+\by \by \right) +\zeta_{s\parallel}^{+} \bz \bz 
\eqno(zetadiag .a)
$$
$$
\ZTA{12} ^{+} =\ZTA{21} ^{+} =
\zeta_{m\perp}^{+} \left(\bx \bx+\by \by \right) +\zeta_{m\parallel}^{+} \bz \bz 
\eqno(zetadiag .b)
$$
where the indices $s$ and $m$ refer to ``self'' and ``mutual'', and the $Oz$
axis has been chosen along $\bR=\bR_1-\bR_2$.

Taking the Brownian limit $M\rightarrow \infty$ for a system with finite $N$,
it is easily found from eqs \gcall{Gam}, \gcall{rel_MM} and \gcall{IPPPP} that the matrix
$\UU^{-1}(s=0)$ appearing in eq \gcall{equality} goes, when $s\rightarrow 0$, to the
finite value
$$
\{\UU^{-1}(s=0)\}_{ab} ={1\over 2}\ \Unity \cdot \left(
2\delta_{ab} -1 \right)
\eqno(IUU_final)
$$
This non-vanishing result can be directly traced back to the conservation
of total (fluid + Brownian) momentum. Consequently we derive from eq. \gcall{equality}
the following relation between the friction tensor matrices $\ZTAN{} $ and
$\ZTA{} ^{+}$, valid for any finite $N$ :
$$
\ZTAN{ab} ={1 \over {k_BT}}\bra \til\FF_a (s=0)  \FF_b\ket= {1 \over 2}
\left(
\ZTA{11} ^{+}-\ZTA{21} ^{+}) \right) \cdot \left(
2\delta_{ab} -1 \right)
\eqno(FF_final)
$$
In other words, keeping in mind eq. \gcall{diff} :
$$
\ZTAN{11} =-\ZTAN{12} =
{\left(\ZTA{11} ^{+}-\ZTA{21} ^{+}) \right)\over 2}
+ {\cal O} \left( {1\over N}\right)
\eqno(FF_final2)
$$ 
We thus have established a first relation between the true friction tensor
and the tensor $\ZTAN{} $ which may be calculated from the total force ACF
of a finite system. From the small-$s$ expansion of $\UU^{-1}(s)$ in eq.
\gcall{equality} one finds that the time integral of the force ACF
$\bra \Ff (t) \Ff (0) \ket$ relaxes {\it exponentially} towards its infinite
time value. The relaxation times associated with the components of the 
fluctuating force perpendicular and parallel to the vector $\bR$ are found in terms of the 
transverse and longitudinal components of the tensors \gcall{zetadiag}, namely
$$
\tau_{\perp/\parallel}^{-1}= {2(\zeta_{s,\perp/\parallel}^{+}+
\zeta_{m,\perp/\parallel}^{+}) \over {Nmk_BT}}
\eqno(relax_time)
$$
These relaxation times are seen to be proportional to the system size. However,
one can easily check that the {\it difference} 
between the time-integral of the self and mutual force ACF's relaxes
towards the difference of the corresponding friction matrices, 
$\ZTA{11} ^{+}-\ZTA{12} ^{+}$, on a much
faster time scale, of the order of the correlation time of the random force.
This fast relaxation is due to a compensation between the slow decays of the separate
functions, and allows accurate estimates of the difference $\ZTA{11} ^{+}-\ZTA{12} ^{+}$
from MD simulation, as illustrated in Fig 1.

At this stage, a second relation is needed to determine the self and mutual friction
coefficients separately from MD simulations of a finite system. In a spirit similar 
to the case of a single Brownian particle \refto{Espanol,Bocquet}, we now consider
the relaxation of the total momentum of $N$ spheres in the presence of two infinitely
massive (i.e. fixed) Brownian spheres. Summing the two Langevin equations 
\gcall{langevin} for $\bP_1$ and $\bP_2$ in the Markovian limit, noting that
$\Fbar_1=-\Fbar_2$, and averaging over a time interval intermediate between the
short time scale of the fluctuating random force and the long time scale associated
with the Brownian particles, we find for the evolution of the total momentum 
$\bP=\bP_1+\bP_2$ :
$$
\dot\bP(t)=-2 \left\{\ZTA{11} + \ZTA{12} \right\}  {\bP(t)\over {mN}}
\eqno(P_t)
$$
Due to the conservation of total momentum, \gcall{P_t} also holds for the momentum
of the fluid $\bP(t)=Nm\bv{} (t)$, where $\bv{} (t)$ denotes the center-of-mass velocity
relative to the fixed Brownian particles. We conclude that the components of the
fluid momentum relax exponentially, on time scales identical to those of the 
integrated ACF's, given in eq. \gcall{relax_time}. In practice, a MD calculation
of the logarithm of the normalized fluid ACF :
$
\log F_\alpha(t)\equiv \log\left[{\bra P_\alpha(t) \cdot P_\alpha(0)\ket_N \over {Nmk_BT}}\right]
$
should be a straight line with slope $2(\zeta_{s,\alpha}^{+}+
\zeta_{m,\alpha}^{+})/ (Nmk_BT)$, where $\alpha = \{\parallel,\perp\}$ 
is either a longitudinal or transverse component. An example of such a plot is shown 
in Fig. 2. The slopes thus provide the second relation between the longitudinal
and transverse elements of the self and mutual friction tensors, which together
with eq. \gcall{FF_final2}, entirely determine the latter.

The various coefficients depend on the distance $R$ between the Brownian particles,
and MD simulations must be carried out for different spacings of the fixed
Brownian spheres. To illustrate the procedure, we quote the values of $\zeta_{s,\alpha}$
and $\zeta_{m,\alpha}$ ($\alpha = \{\parallel,\perp\}$) obtained for a size ratio
of the Brownian and fluid spheres $\Sigma/\sigma=2$, and a distance $R/\Sigma=1$.
The packing fraction of the fluid is $\eta=0.246$. The following values are extracted from plots as shown in  Figs. 1 and 2 :
$\zeta_{s,\parallel}^*= \zeta_{s,\parallel}/m\nu_{mM}= 12.1$,
$\zeta_{s,\perp}^*= \zeta_{s,\perp}/m\nu_{mM}= 6.7$,
$\zeta_{m,\parallel}^*= \zeta_{m,\parallel}/m\nu_{mM}= -0.3$,
$\zeta_{m,\perp}^*= \zeta_{m,\perp}/m\nu_{mM}= -0.3$, where $\nu_{mM}$
is the collision frequency between the bath and one Brownian particle.
Contrary to the prediction of hydrodynamics, the value remains finite when
the two Brownian particles are at contact. In general our numerical results
indicate strong deviations on these scales from the $R$-dependence of the friction coefficient
from hydrodynamics.
More complete report, for various size ratios $\Sigma/\sigma$ and 
distances $R$ will be published elsewhere, and confronted with the predictions
of macroscopic hydroynamics.

\refis{Piasecki} J. Piasecki, L. Bocquet, J.P. Hansen, Physica A {\bf 218},
125 (1995).

\refis{Beenaker} see e.g. C.W.J. Beenakker anf P. Mazur, Physica A
{\bf 131} 311 (1985) and references therein

\refis{Deutch} J.M. Deutch and I. Oppenheim, J. Chem. Phys. {\bf 54},
3541 (1971).

\refis{Murphy} T.J. Murphy and J.L. Aguirre, J.Chem. Phys. {\bf 57}, 2098
(1972).

\refis{Espanol} P. Espa\~nol, I. Z\'u\~niga,
 J. Chem. Phys. {\bf 98} 574 (1993); 

\refis{Cukier} R.I. Cukier and J.M. Deutch, Phys. Rev {\bf
177}, 240 (1969).

\refis{Bocquet} L. Bocquet, J. Piasecki and J.P. Hansen, J. Stat.
Phys., {\bf 76} 505 and 527 (1994).

\refis{Kirkwood} J. Kirkwood, J. Chem. Phys. {\bf 14}, 180 (1946).

\refis{Israel} J.N. Israelachvili, ``Intermolecular and Surface Forces''
(Academic press, London, 1990).

\refis{Mazo} R.M. Mazo, J. Stat. Phys. {\bf 1}, 559 (1969).

\refis{Asakura} S. Asakura and F. Oosawa, J. Polymer Sci. {\bf 33},
183 (1958); A. Vrij, Pure and Appl. Chem. {\bf 48}, 471 (1976).

\refis{Alley} W.E. Alley, PhD. Thesis, Lawrence Livermore Laboratory
report UCRL-52815 (1979).

\references

\endreferences

\figurecaptions

{\bf Figure 1} : Time-dependence of the reduced self (upper curve) and mutual (lower curve) friction coefficient, with appropriate sign ($\epsilon_{ab}=+1$ if $a=b$ and $-1$ if $a\ne b$). $\zeta_{ab}^{*}(t)=\zeta_{ab}(t)/m\nu_{mM}$ is
defined as the time-integral of the corresponding force ACF, up to time t
(see eq. (2)), with $\nu_{mM}$ the fluid-Brownian particle collision frequency.
The half difference
between the two previous estimates (open circles), is shown to 
converge towards the same asymptotic value on a much faster time-scale, of the order of a few
collision time inside the fluid.

{\bf Figure 2} : Logarithmic plot of the normalized ACF, $F_\alpha(t)$, of the
total fluid momentum versus reduced time $\nu_{mM} t$. The solid (resp. dashed) line corresponds to the
component of the momemtum ACF perpendicular (resp. parallel) to $\bR=\bR_1-\bR_2$.

\endfigurecaptions

\bye